\documentstyle[buckow1,12pt,epsf]{article}

\begin {document}

\def\titleline{
   \begin{flushleft}
      \vspace{-3.50cm}
      {\normalsize HUB--EP--98/2}   \\[-0.2cm]             
      {\normalsize FU--HEP/98-2}    \\[-0.2cm]             
      {\normalsize January 1998}    \\                     
      \vspace{2.00cm}                                      
   \end{flushleft}

$O(a)$ Improvement for Quenched Wilson Fermions
}
\def\authors{
\begin{flushleft}
S.~Capitani\1ad, M.~G\"ockeler\2ad, R.~Horsley%
\footnote{Talk presented by R. Horsley at the 31st Ahrenshoop
Symposium on the Theory of Elementary Particles,
September 1997, Buckow, Germany.}\3ad,
B. Klaus\4ad$^,$\5ad,  H.~Oelrich\4ad, H.~Perlt\6ad,
D. Petters\4ad$^,$\5ad, D.~Pleiter\4ad$^,$\5ad, P.~Rakow\4ad,
G.~Schierholz\1ad$^,$\4ad, A.~Schiller\6ad and P.~Stephenson\7ad
\end{flushleft}
}
\def\addresses{
\begin{footnotesize}
\begin{flushleft}
\1ad
Deutsches Elektronen-Synchrotron DESY, D-22603 Hamburg, Germany \\
\2ad
Institut f\"ur Theoretische Physik, Universit\"at Regensburg,
D-93040 Regensburg, Germany \\
\3ad
Institut f\"ur Physik, Humboldt-Universit\"at zu Berlin,
D-10115 Berlin, Germany \\
\4ad
DESY, HLRZ und Institut f\"ur Hochenergiephysik, D-15735 Zeuthen, Germany \\
\5ad
Institut f\"ur Theoretische Physik,
Freie Universit\"at Berlin, D-14195 Berlin, Germany  \\
\6ad
Institut f\"ur Theoretische Physik, Universit\"at Leipzig,
D-04109 Leipzig, Germany  \\
\7ad
Dipartimento di Fisica, Universit\`a degli Studi di Pisa
e INFN, Sezione di Pisa, 56100 Pisa, Italy
\end{flushleft}
\end{footnotesize}
}
\def\abstracttext{
\begin{footnotesize}
We briefly describe some of our recent results for the mass spectrum
and matrix elements using $O(a)$ improved fermions
for quenched {\it QCD}. Where possible a comparison is made
between improved and Wilson fermions.
\end{footnotesize}
}
\makefront

\section{Introduction}

Upon discretising the {\it QCD}-Lagrangian in the Wilson formulation
we find that the gluon piece has $O(a^2)$ errors, but for the
fermion part, the additional Wilson term, necessary to avoid the fermion 
`doubling' problem gives us $O(a)$ errors.
Thus we would expect that looking at any physical quantity,
for example a mass ratio, we have
\begin{equation}
   {m_H\over m_{H^\prime}} = r_0 + a r_1 + a^2 r_2 + O(a^3).
\label{physical_wilson}
\end{equation}
Symanzik developed a systematic perturbative progamme to $O(a^n)$
in which a basis of irrelevant operators is added to the Lagrangian
to completely remove $O(a^{n-1})$ effects. Restricting this to
on-shell (or physical) quantities, \cite{luescher85a},
enables the equations of motion
to be used to reduce the required set of operators, in both
the action and for improved operators in matrix elements.
In this talk we briefly report our progress
on computing the hadron spectrum and matrix elements for $O(a)$ improved
fermions, where in this case we expect
for a physical quantity such as a mass ratio
\begin{equation}
   {m_H\over m_{H^\prime}} = r_0 + a^2 r_2^\prime + O(a^3).
\label{physical_magic}
\end{equation}
The ground-work has been laid by the Alpha collaboration
who succeeded in non-perturbatively calculating several improvement
coefficients for $\beta\equiv 6.0/g^2 \ge 6.0$.
Reviews and further references may be found in \cite{sint96a}.
We shall here neither describe their method
nor go into details of how the mass spectrum or matrix elements are
computed on the lattice. A description of our spectrum results is
given in ref.~\cite{goeckeler97a} while for Wilson matrix elements
see, for example, \cite{goeckeler95a}.
Up to now we have generated configurations on $(16^3,24^3)\times32$
lattices at $\beta = 6.0$ and on $24^3\times 48$
lattices at $\beta =6.2$. We have used the string tension, $\sqrt{K}$,
to determine the lattice spacing $a$ at these two $g^2$ values.
A physical value of $\sqrt{K}=0.427\mbox{GeV}$ is taken to set the scale.

\section{The mass spectrum}

\begin{figure}[h]
   \begin{tabular}{cc}
      \hspace{-1.0cm}
      \epsfxsize=8.00cm \epsfbox{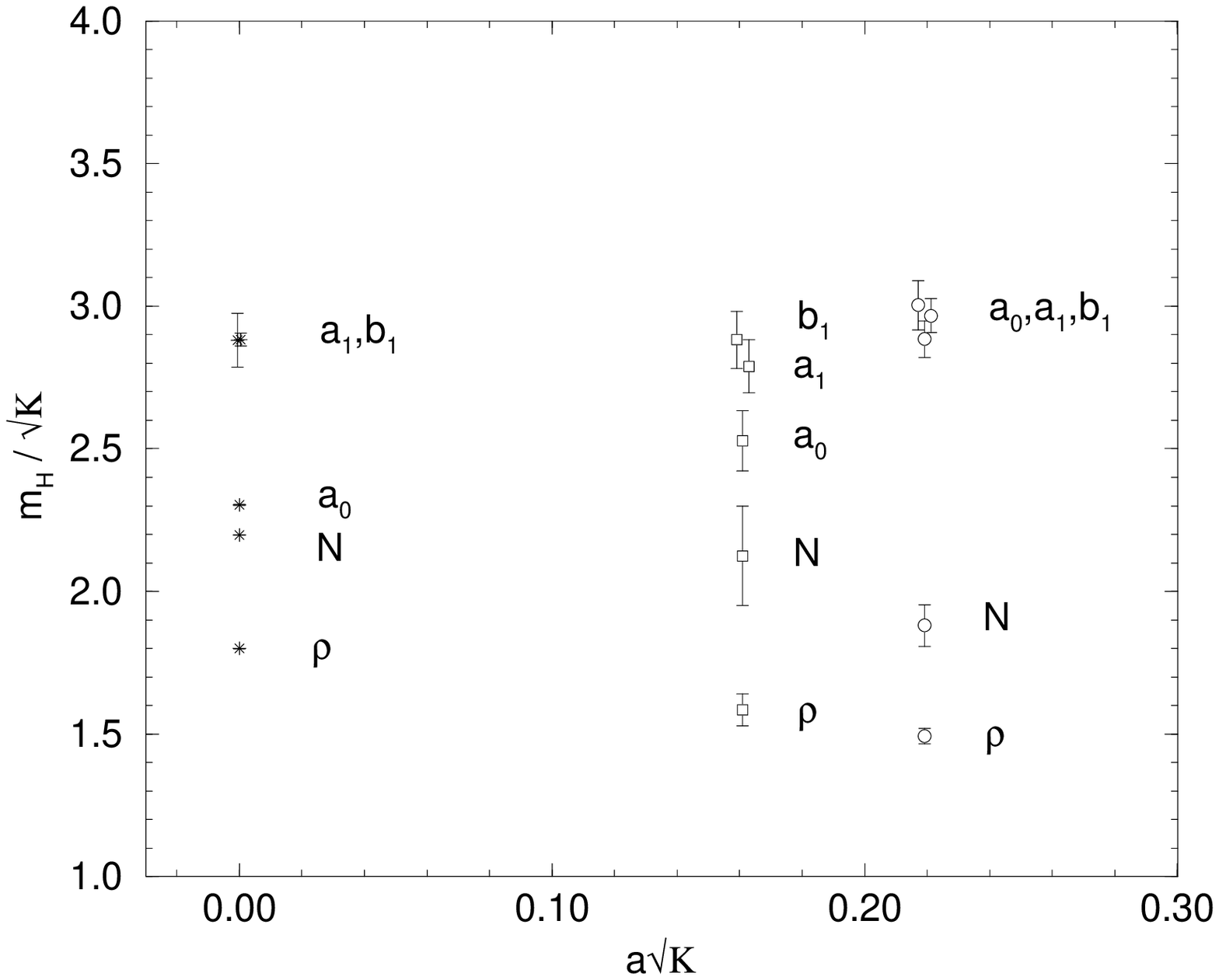}  &
      \hspace{-1.0cm}
      \epsfxsize=8.00cm \epsfbox{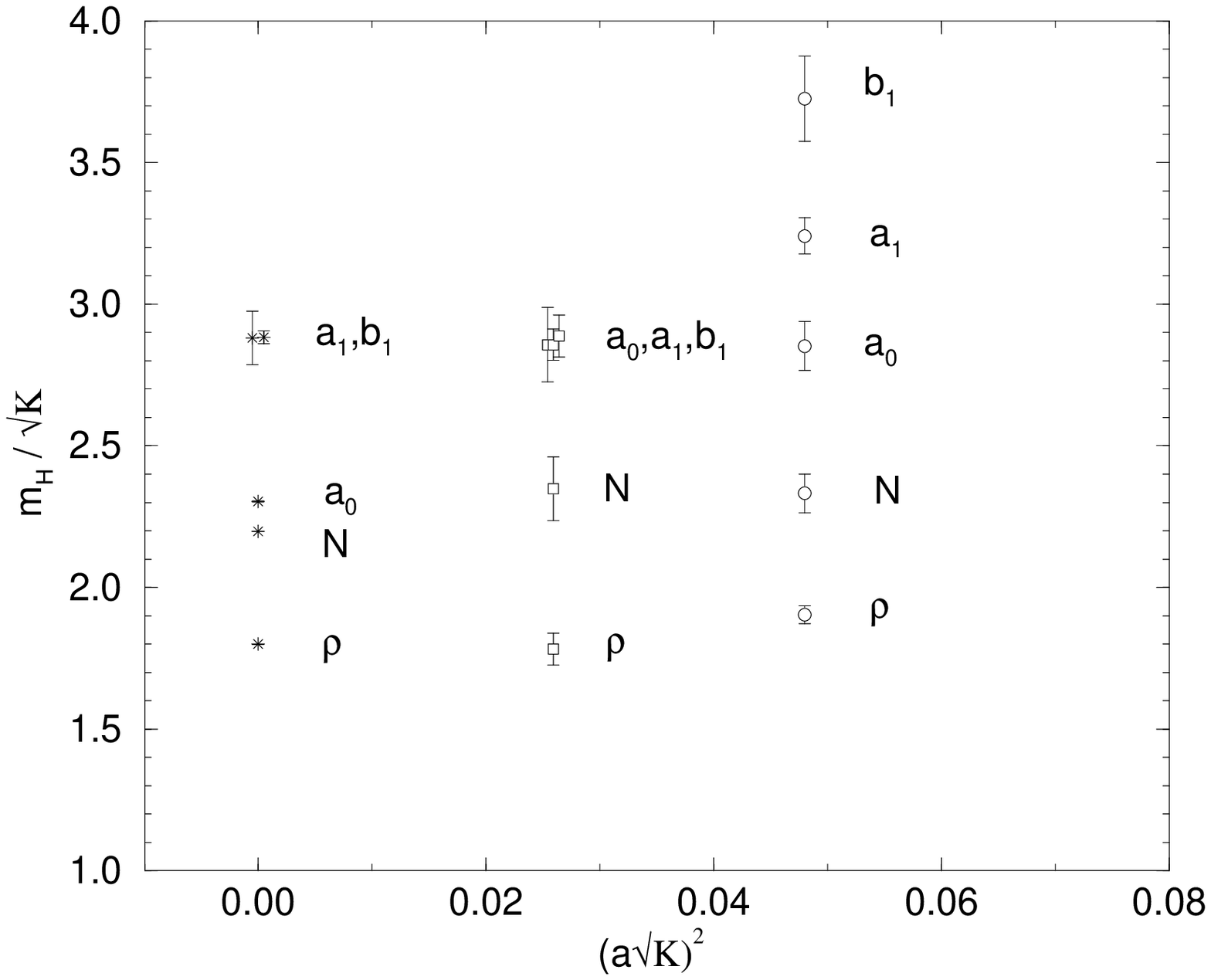}
   \end{tabular}
   \caption{\footnotesize
            {The hadron spectrum for light hadrons at $\beta = 6.0$ (circles),
            $6.2$ (squares) using Wilson fermions (left picture)
            and using $O(a)$ improved fermions (right picture)
            against the lattice spacing, $a$. Experimental numbers
            are shown as stars.}}
   \label{fig_hadron}
\end{figure}

\begin{figure}[h]
   \hspace{3.0cm}
   \epsfxsize=8.00cm \epsfbox{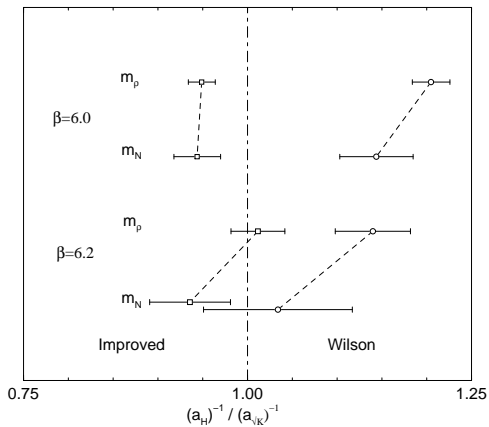}
   \hspace{-1.0cm}
   \caption{\footnotesize{
            $a_H^{-1}$ estimates using $H = \rho$ or $N$ as a scale
            normalised to the string tension for Wilson 
            fermions (circles) compared to $O(a)$ improved masses
            (squares).}}
   \label{fig_a}
\end{figure}

We have computed the $\rho$ ($J^{PC}= 1^{--}$) and nucleon ($N$) masses
together with the $a_0$ ($0^{++}$), $a_1$ ($1^{++}$) and
$b_1$ ($1^{+-}$) masses. The latter three mesons being
{\it p}-wave states are difficult to measure with our symmetric
sources and our results should only be taken as 
indicative of possible trends. In Fig.~\ref{fig_hadron}
we plot this hadron spectrum both for Wilson and $O(a)$
improved fermions. Roughly the same amount of {\it CPU} time has gone
into producing each of these pictures.
From eqs.~(\ref{physical_wilson},\ref{physical_magic})
we expect that the dominant discretisation
terms are such that linear extrapolations in $a$ for Wilson fermions
and $a^2$ for $O(a)$ improved fermions are sufficient.
Clearly with only two points we must
limit ourselves here to qualitative observations%
\footnote{Presently, for $O(a)$ improved fermions, most groups have
concentrated on $\beta = 6.0$, $6.2$, \cite{mendes97a},
but there has been a recent interesting attempt to go to
lower $\beta$ values, \cite{edwards97a}.}.
First we note that due to the absence of the $O(a)$
term for the improved fermions, the convergence to the continuum
limit is faster. On the other hand it does appear that,
especially for heavier particles, the signal fluctuates more.
It also seems that at least for the $\rho$ and $N$ particles,
$O(a)$ improved fermions results lie closer to the continuum result --
the Wilson results have a noticeable gradient. 
An alternative way of looking at these results is given in
Fig.~\ref{fig_a}. While it is not necessary that quenched {\it QCD}
should reproduce exactly the continuum spectrum, in many present-day
applications a hadron mass (typically the $\rho$ or $N$) is used to set
the scale. Potential discrepancies between these scales can be revealed
by normalising them against the string tension
(which has $O(a^2)$ errors). Again this confirms the previous impression: 
$O(a)$ improved fermions for the $\rho$ or $N$ perform better than
Wilson fermions.

\section{Matrix elements}

\begin{figure}[h]
   \begin{tabular}{cc}
      \hspace{-1.0cm}
      \epsfxsize=8.00cm \epsfbox{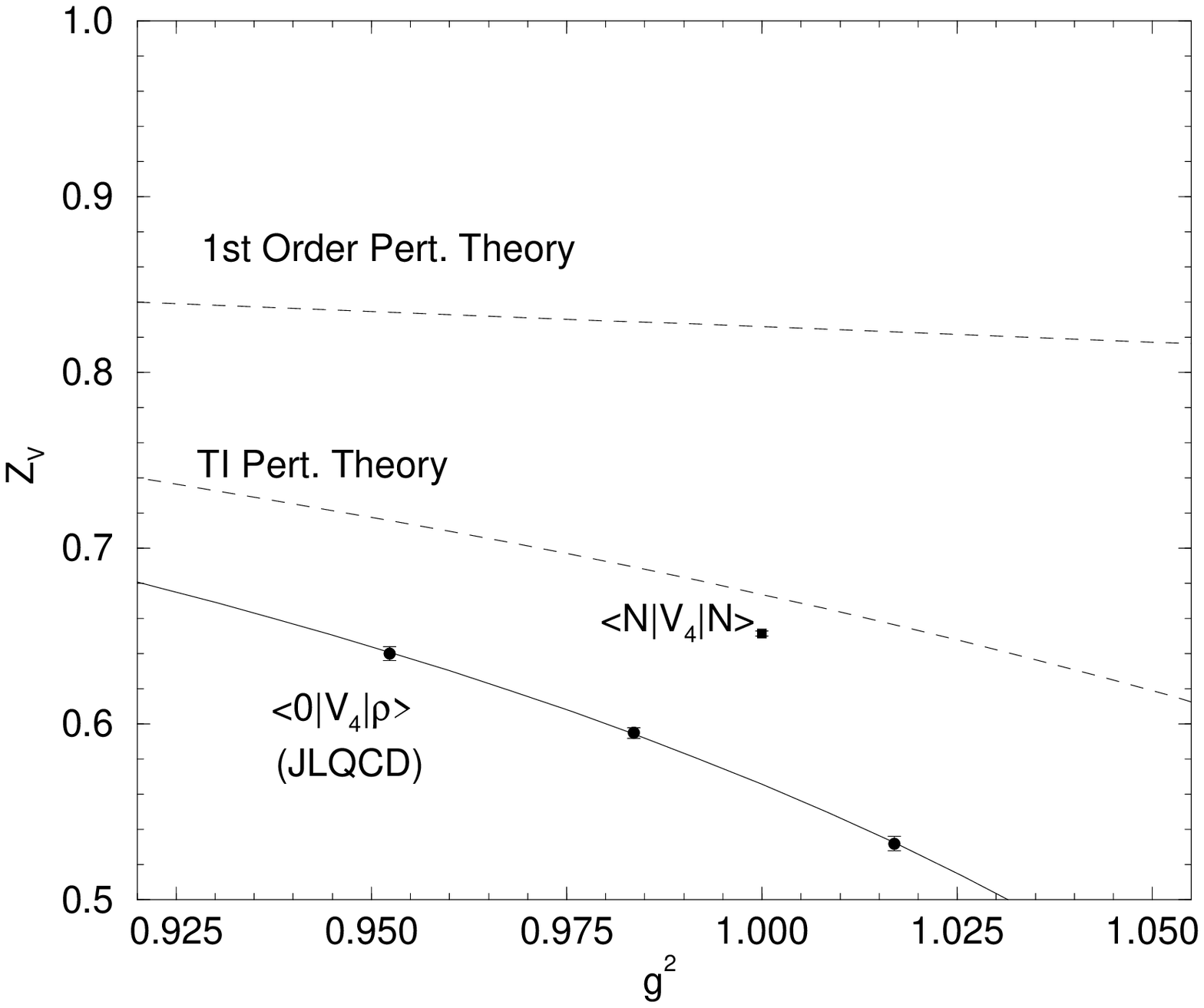}  &
      \hspace{-1.0cm}
      \epsfxsize=8.00cm \epsfbox{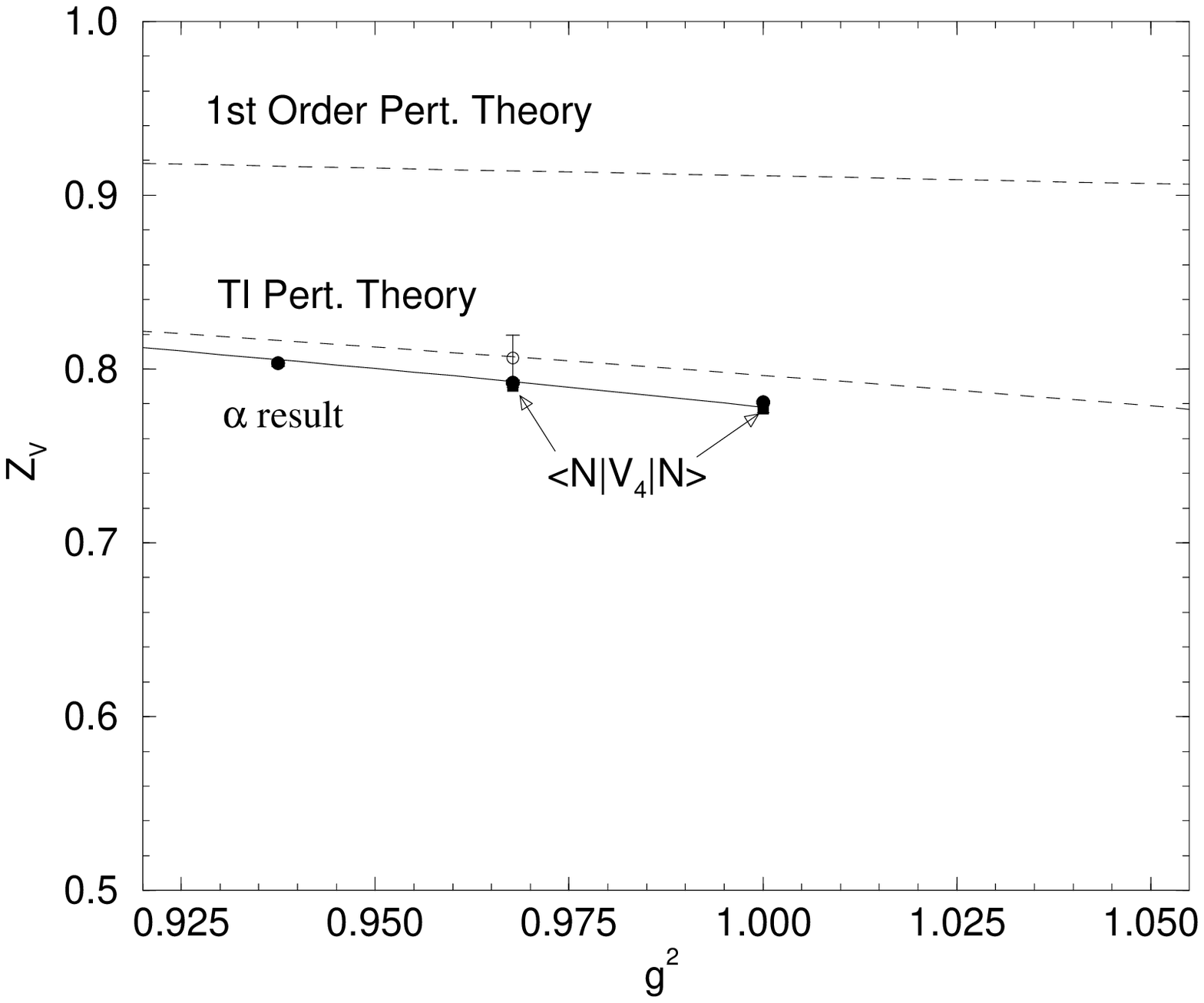}   \\
      \hspace{-1.0cm}
      \epsfxsize=8.00cm \epsfbox{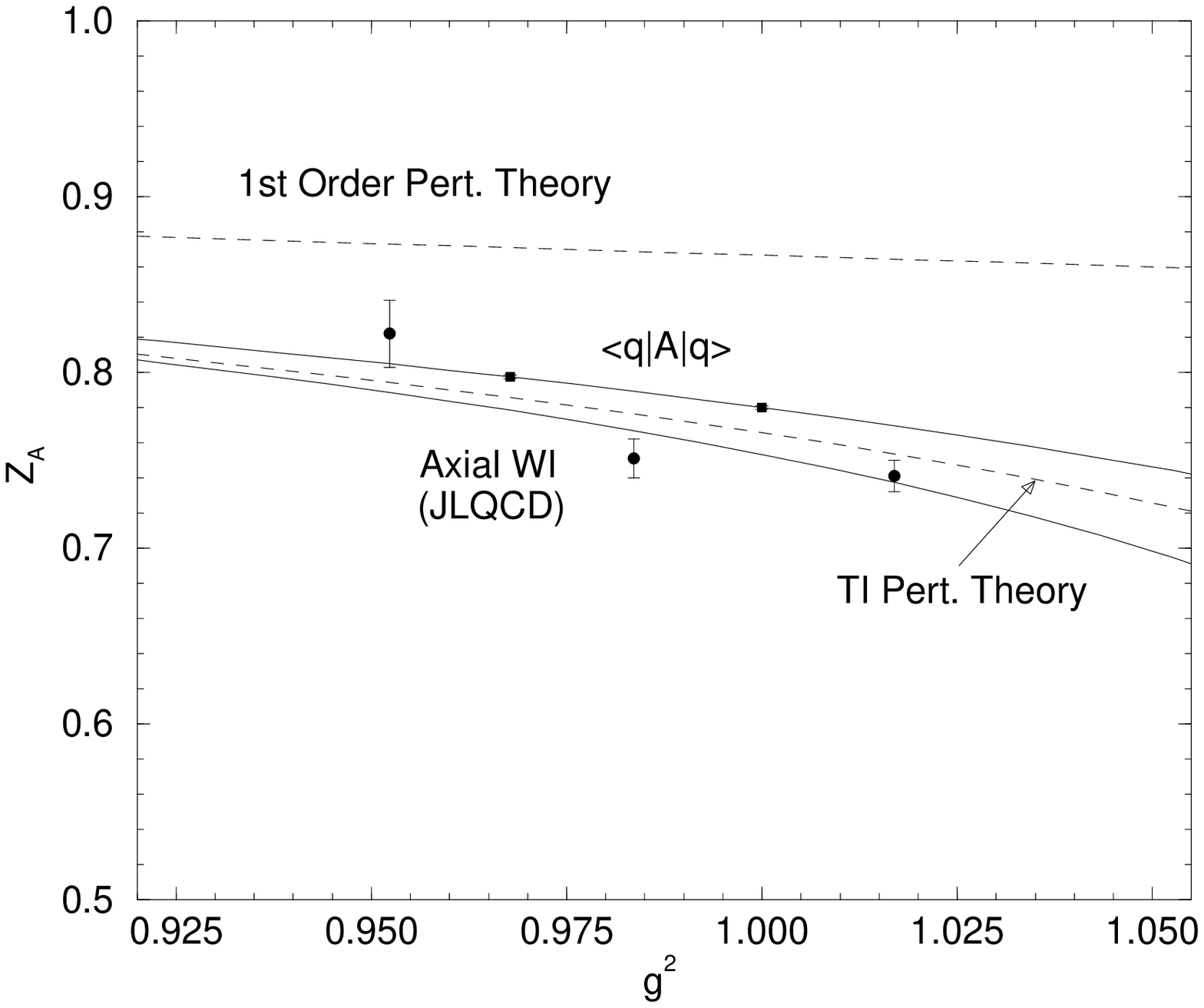}  &
      \hspace{-1.0cm}
      \epsfxsize=8.00cm \epsfbox{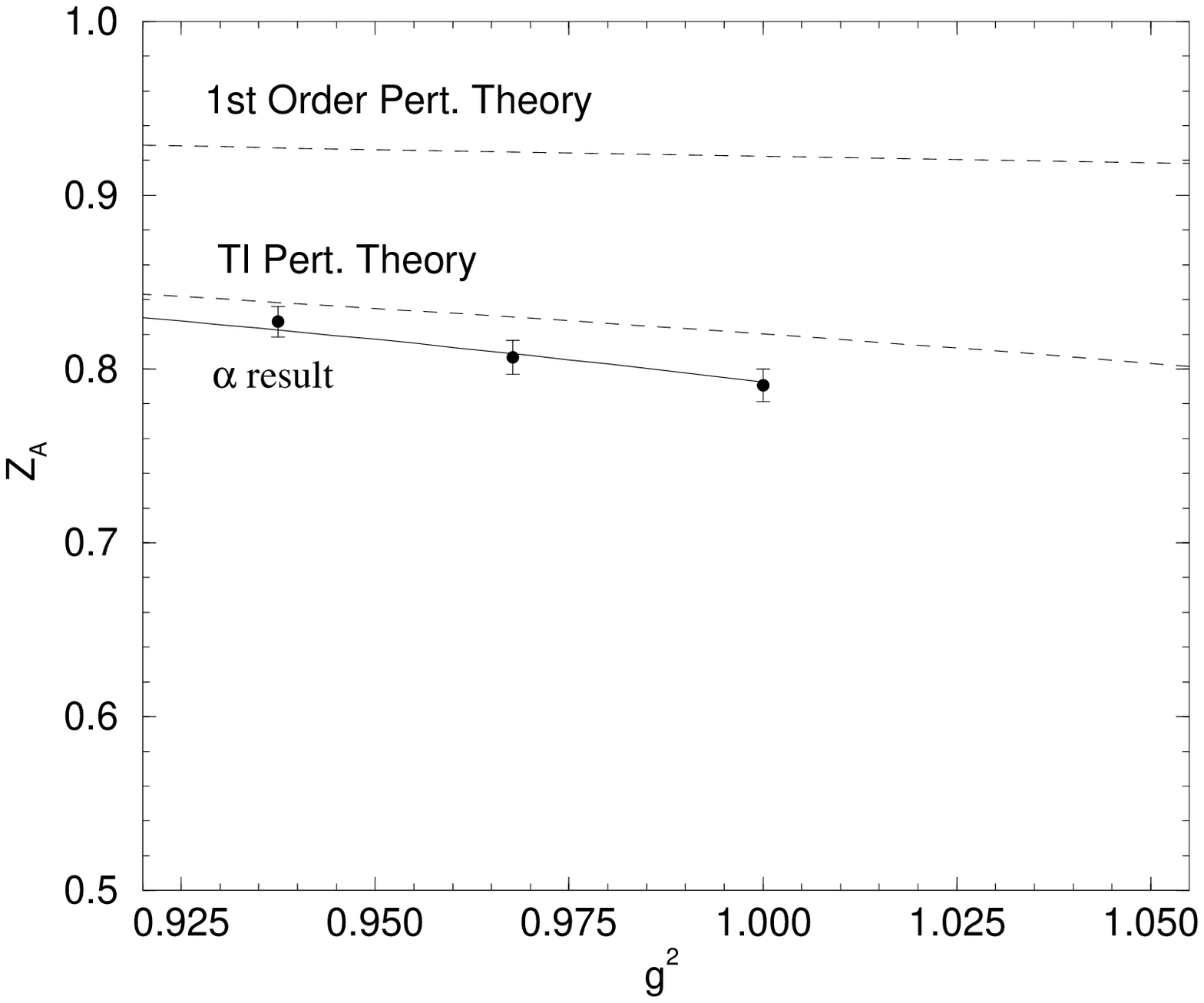}
   \end{tabular}
   \caption{\footnotesize{
            Various determinations of $Z_V$ (upper row)
            and $Z_A$ (lower row) for Wilson (left column)
            and $O(a)$ improved fermions (right column).
            The tadpole improved ({\it TI}) results have been obtained
            from first order perturbation results, 
            $Z_O(g^2) = 1 + c_O g^2 + O(g^4)$, \cite{capitani97b},
            using the procedure given in \cite{goeckeler97a}.
            Also plotted are results using the local vector current
            between nucleon, \cite{capitani97a} (filled squares)
            and $\rho$ states, \cite{aoki97a} (filled circles),
            and the local axial current between quark states,
            \cite{goeckeler97b} (filled squares). Where appropriate,
            a simple Pad\'e interpolation/extrapolation has been applied,
            $Z_O(g^2) = (1 + p_Og^2 + q_Og^4)/(1+r_O g^2)$
            with $p_O-r_O = c_O$. For the $O(a)$ improved
            fermions we also give the Alpha results (filled circles)
            and Pad\'e extrapolation, \cite{luescher96a} and from
            \cite{divitiis97a} (open circle).}}
   \label{fig_zv+za}
\end{figure}

\begin{figure}[t]
   \begin{tabular}{cc}
      \hspace{-1.0cm}
      \epsfxsize=8.00cm \epsfbox{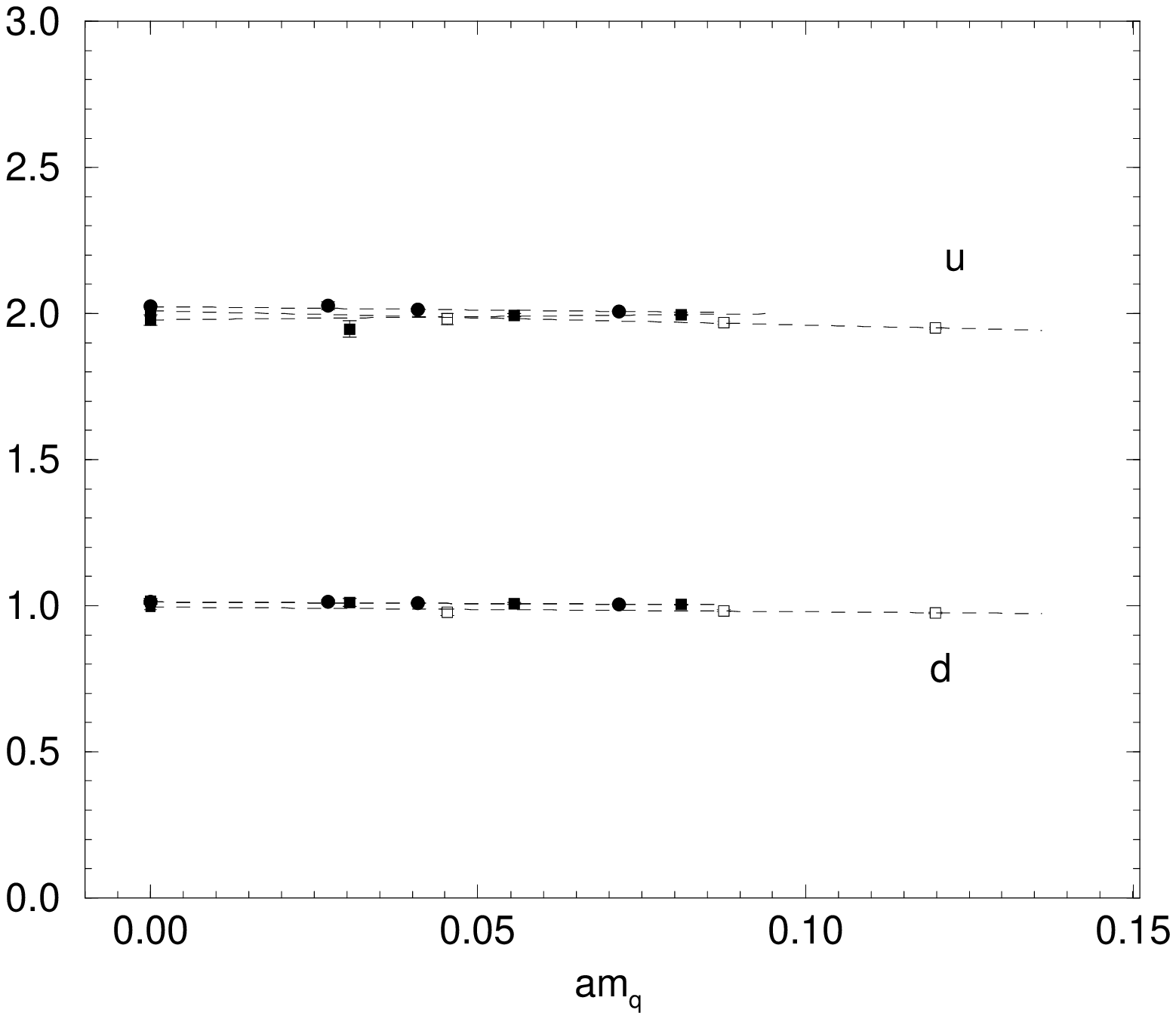}  &
      \hspace{-1.0cm}
      \epsfxsize=8.00cm \epsfbox{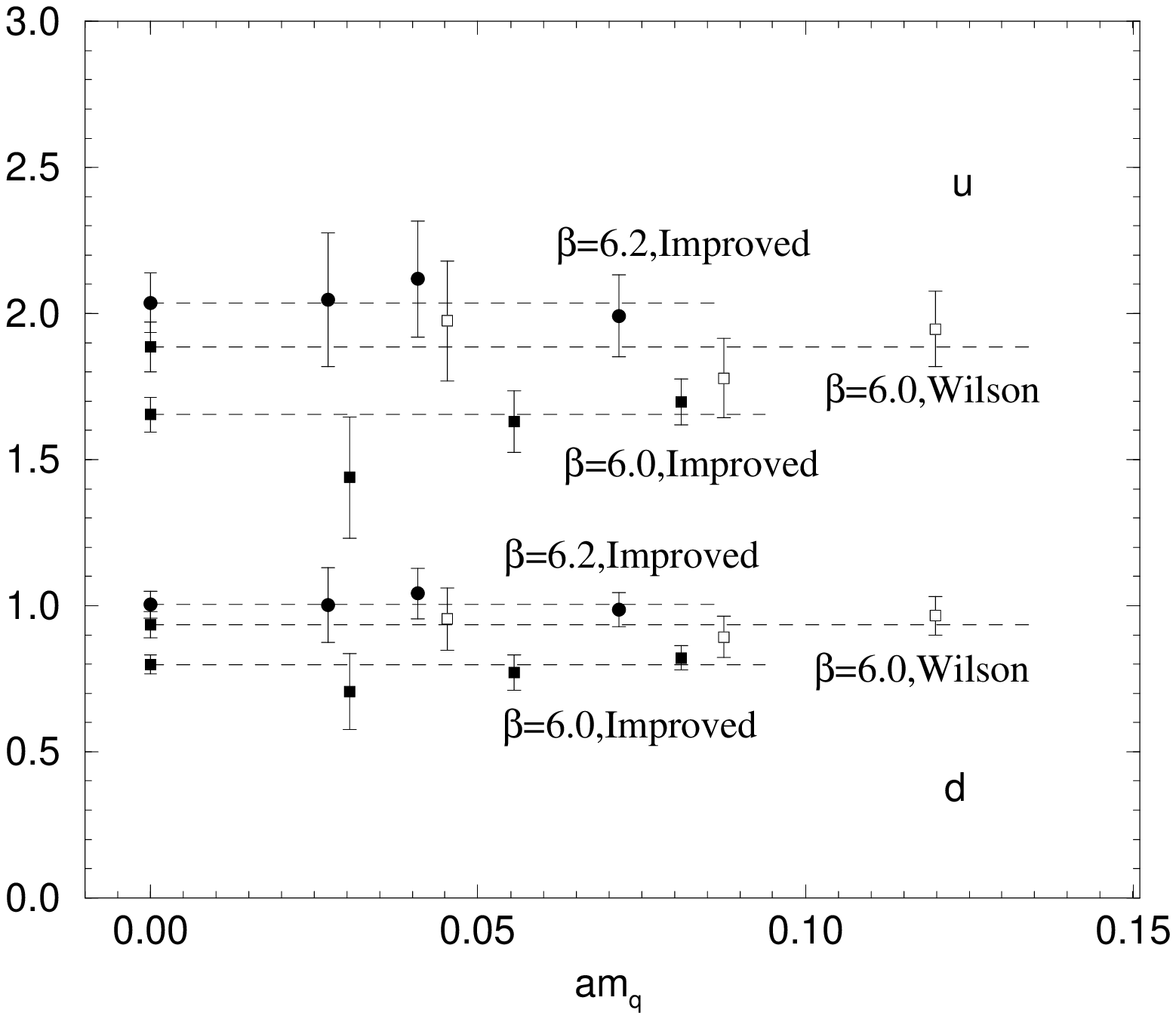}
   \end{tabular}
   \caption{\footnotesize{
            $\langle N|{\cal V}^R_4|N\rangle$, left picture,
            $\langle N|{\cal V}^R_1 |N\rangle \times E_N/p_1$,
            right picture, at $\vec{p}=(p_1,0,0)$ with $p_1$
            being the minimum possible momenta,
            namely $p_1 = 2\pi/16a \approx 765\mbox{MeV}$
            or $2\pi/24a \approx 694\mbox{MeV}$ at $\beta = 6.0$, $6.2$
            respectively. The expected values are $\chi_u=2$, $\chi_d=1$.
            Filled circles, squares denote $O(a)$
            improved fermions at $\beta=6.2$, $6.0$ respectively,
            while open squares are for $\beta=6.0$ Wilson fermions.
            To guide the eye, constant fits are made to the chiral limit
            for the right hand picture.}}
   \label{fig_V0+V1}
\end{figure}

Matrix elements, such as decay constants or moments of structure
functions, are more complicated to calculate than masses as the operators
must be appropriately renormalised, so that
${\cal O}^R = Z_O (1+b_O am_q) {\cal O}$,
with ${\cal O} = O + \sum_I c_I aO_I$.
For present-day $\beta$ values, non-perturbative ({\it NP}) determinations
of $Z_O$, $b_O$ and $\{ c_I \}$ are preferable.
(Here we are mainly interested in matrix elements in the chiral
limit, so we do not need $b_O$; the other irrelevant 
operators $O_I$ are only required in the improved case to ensure
complete $O(a)$ cancellation.) In Fig.~\ref{fig_zv+za} we show
$Z_V$ ($V_\mu = \overline{q}\gamma_\mu q$) and
$Z_A$ ($A_\mu = \overline{q}\gamma_\mu\gamma_5 q$)
for both Wilson and $O(a)$ improved fermions. We see that in all 
cases, while first order perturbation theory lies $\ge 10\%$ away
from {\it NP} computations, tadpole improvement ({\it TI}) always gives
results closer to the {\it NP} line. For the Wilson case we expect to find
larger discrepancies (of $O(a)$) between the various {\it NP} estimates
of $Z$ than for improved fermions (of $O(a^2)$). This seems most
obvious for $Z_V$. We next note that both the Alpha, \cite{luescher96a},
and our non-perturbative determination of $Z_V$ for $O(a)$ 
improved fermions are in very good agreement with each other.
From current conservation we expect
$\langle N|{\cal V}^R_\mu |N\rangle = p_\mu/E_N \chi_q$
with $\chi_u = 2$, $\chi_d = 1$.
This at (eg) zero momentum allows a non-perturbative
determination of $Z_V$ (and $b_V$), for both Wilson and $O(a)$ improved
fermions. This result is exemplified in Fig.~\ref{fig_V0+V1} where we perform
a check of momentum effects and restoration of rotational invariance,
by computing $\langle N|{\cal V}^R_4|N \rangle$,
$\langle N|{\cal V}^R_1|N \rangle \times E_N/p_1$.
While all results for $\langle N|{\cal V}^R_4|N\rangle$ are consistent
with each other, $\langle N|{\cal V}^R_1 |N\rangle \times E_N/p_1$
shows some spread; indeed at $\beta=6.0$, Wilson fermions seem
to perform better than the $O(a)$ improved fermions.

\begin{figure}[h]
   \begin{tabular}{cc}
      \hspace{-1.0cm}
      \epsfxsize=8.00cm \epsfbox{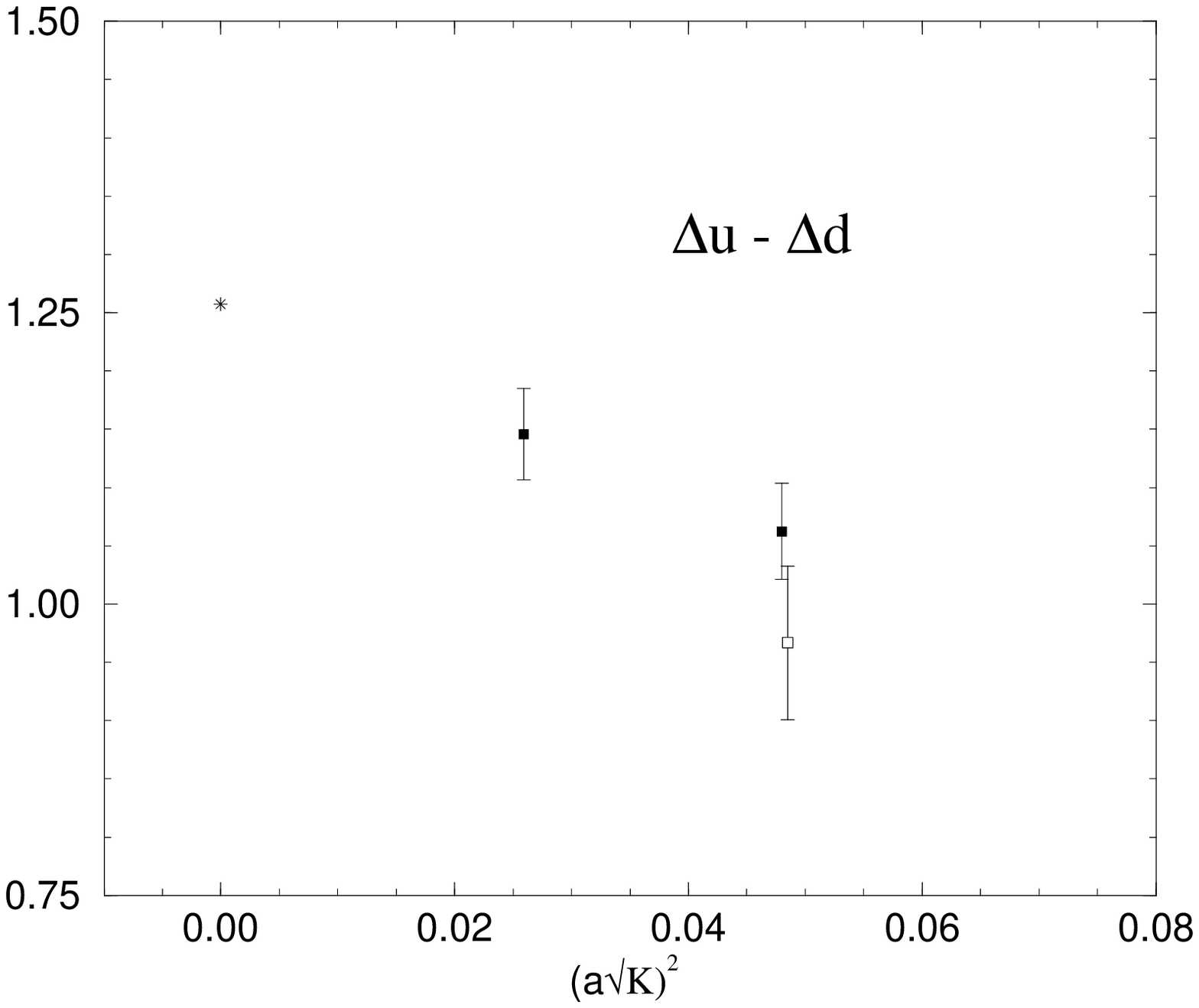}  &
      \hspace{-1.0cm}
      \epsfxsize=8.00cm \epsfbox{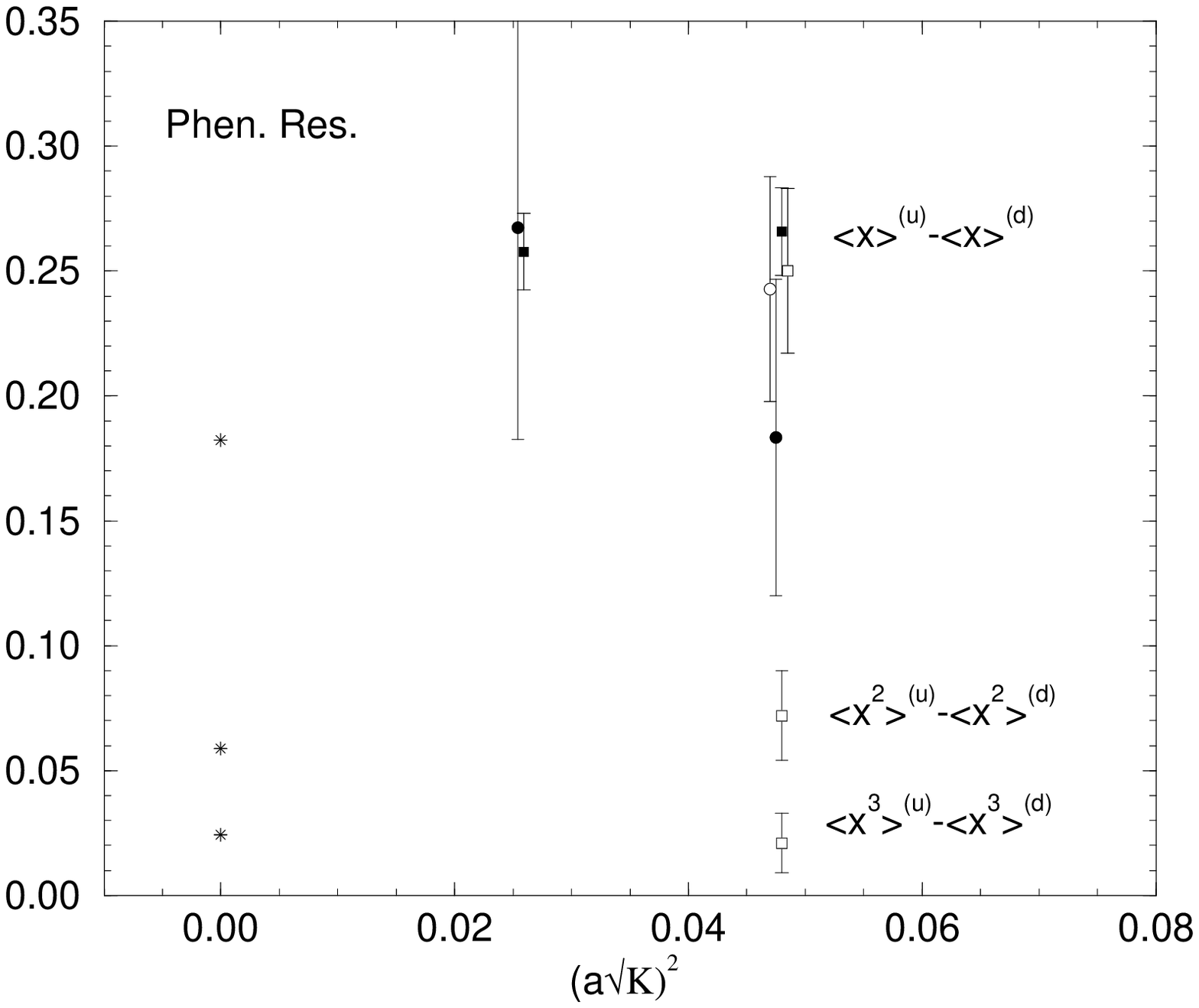}
   \end{tabular}
   \caption{\footnotesize{
            $g_A = \Delta u - \Delta d$, left picture, and
            $\langle x^n \rangle^{(u)} - \langle x^n \rangle^{(d)}$
            moments at a scale of $\mu \sim 1.95\mbox{GeV}$,
            right picture. The phenomenological results are
            denoted with stars ($g_A =1.26$ and 
            $\langle x^n \rangle^{(u)} - \langle x^n \rangle^{(d)}$
            by the {\it MRS} parametrisation, \cite{martin95a}).
            $O(a)$ improved fermion results are given by 
            filled symbols. For $\langle x \rangle$, results
            using a zero-momentum diagonal matrix element
            are given by a square while the non-zero momentum
            off-diagonal results are denoted with a circle.}}
   \label{fig_ga+xn}
\end{figure}

In the left picture of Fig.~\ref{fig_ga+xn} we apply the
Alpha result for $Z_A$ to a determination of $g_A$.
There seems to be an approach to
the experimental value of $g_A$ (but with rather large $O(a^2)$
corrections). For the moments of the quark distributions
in the nucleon, we obtain the results shown in the right hand picture in
Fig.~\ref{fig_ga+xn}. The lowest moment, $\langle x \rangle$,
derived from the operator $\overline{q}\gamma_\mu D_\nu q$
is interesting, not only because it contains one derivative,
but also because there are two distinct lattice representations
involving diagonal or off-diagonal elements, the latter requiring
for its evaluation a non-zero spatial momentum.
(Again we always used the minimum possible.)
This potentially allows a better
study of the approach to the continuum limit. At present a 
non-perturbative renormalisation of $\langle x \rangle$ is not known;
however no tadpole term contributes to first order perturbation theory,
so that this automatically lies close to the {\it TI} result,
so we might hope that any error in the renormalisation constant is small.
A further problem for the $O(a)$ improved operators is the addition
of at least one irrelevant operator. However, as discussed in
\cite{capitani97a}, we hope that its effect is also small.
For $\langle x \rangle$, from Fig.~\ref{fig_ga+xn},
all results appear to be reasonably consistent with each other,
and for $O(a)$ improved fermions there does not seem to
be any large $O(a^2)$ effect. However, there seems to be a larger
deterioration of the signal (in comparison to the Wilson case) when
introducing a momentum into the operator.

In conclusion, it would seem that $O(a)$ Symanzik improvement does
indeed bring us closer to the goal of calculating continuum masses
and matrix elements in {\it QCD}.
The numerical calculations were performed on the Quadrics facility
at {\it DESY-IfH}.

\end{document}